\documentclass[12pt]{iopart}

\usepackage{graphicx}

\begin{document}

\title[How far away is infinity?]{How far away is infinity? An
  electromagnetics exercise to develop intuition regarding models}

\author{Álvaro Suárez$^1$, Martín Monteiro$^2$, Mateo Dutra$^3$,
  Arturo C. Martí $^3$}

\address{$^1$CFE-ANEP, Uruguay; $^2$Universidad ORT Uruguay;
  $^3$Universidad de la República, Uruguay}
\ead{$^a$alsua2000@gmail.com, $^b$fisica.martin@gmail.com,
  $^c$mateodutrafisica@gmail.com,$^d$marti@fisica.edu.uy}
\vspace{10pt}
\begin{indented}
\item[]Abril 2021
\end{indented}

\begin{abstract}
The estimation of the electric field in simple situations provides an
opportunity to develop intuition about the models used in physics. We
propose an activity aimed at university students where the electric field of a finite line of charge is compared, analytically or numerically, with the fields of an infinite line and
of a point charge. Contrary to intuition, it is not necessary to get
very close for the line charge to be considered infinite, nor to move
very far away for the finite line field to resemble that of a point
charge. We conducted this activity with a group of students and found
that many of them have not yet developed an adequate intuition about
the approximations used in electromagnetism.
\end{abstract}

%
%
%
%
%

\noindent
\textbf{Introduction}
Modeling in physics is based on subtle approximations. When teaching
the subject, however, the limits of its validity are not always
explicitly stated or emphasized. Although teaching strategies based on
model building have been developed
\cite{brewe2008modeling,pawl2009modeling} with proven success,
modeling is generally not given due emphasis in traditional
courses. Point particles, plane waves and rigid bodies are some of the
concepts proposed to approximate reality in different situations, but
they are often used superficially. In addition to the students' own
difficulties in understanding and using the different models
\cite{etkina2006role}, the lack of proper discussion may lead them to
become unmotivated and conceive physics as being far from reality
\cite{redish2020using}.

Teaching electromagnetism at introductory levels does not escape the
aforementioned difficulties, involving the frequent use of several
idealized concepts such as point charges, dipoles, and infinite solenoids. Additionally, the application of some of the laws
of electromagnetism, especially Gauss' law, requires the consideration of abstract objects, Gaussian surfaces, to apply this law and find expressions for the fields with relative ease. It is also common to make different approximations to obtain expressions of the fields at small or large distances from the sources. In order to develop intuition regarding these concepts, we propose the following exercise.

\begin{figure}
    \centering
    \includegraphics[width = 0.33\columnwidth]{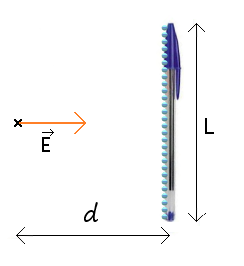}
    \caption{The situation presented to the students focuses on the
      estimation of the electric field at a distance d from a pen
      charged by rubbing.}
    \label{fig1}
\end{figure}

\noindent
\textbf{The field of a finite  line of charge.}
Imagine an ordinary elongated insulating object, e.g., a ballpoint
pen, which is rubbed against the hair and acquires a negative charge
that is distributed more or less uniformly, as shown in Fig.~\ref{fig1}. If we consider a nearby point, the electric field will be
approximately equal to that of an infinite line with the same linear
charge density. The question that naturally arises is how far away can we place ourselves so that the two electric fields differ by less than a certain amount. In the opposite scenario, if we consider a distant point, the electric field will be similar to that of a point charge of equal magnitude. Then what is the minimum distance for the two fields to differ by less than a certain amount? As we will see below, the answers are surprising to most of our students.

We consider the electric field of a bar of length $L$ with a uniformly distributed charge. The magnitude of the field at a distance $d$ on a
perpendicular line passing through its midpoint is
\begin{equation*}
    E_{bar} = \frac{q}{4\pi\varepsilon_0d^2}\frac{1}{\sqrt{\frac{L^2}{4d^2}+1}},
\end{equation*}
the quotient of the two fields results in
\begin{equation*}
    \frac{E_{bar}}{E_q} = \frac{1}{\sqrt{\big(\frac{L}{2d}\big)^2+1}}.
\end{equation*}
As noted above, for distant points tends both fields tend to have the same value. A possible criterion to consider the fields to be approximately equal is that they differ by
less than 5\%, i.e., ${E_{bar}}/{E_q}>0.95$. It is easy to see
that is condition is met if $d>1.52L$, or that the distance must be
greater than one and a half times the length of the bar.

Regarding the comparison between the field of the bar with an infinite line, the electric field needs to be expressed as a function of the linear density of  charge $\lambda=Q/L$,
\begin{equation*}
    E_{bar} = \frac{\lambda}{2\pi\varepsilon_0d}\frac{1}{\sqrt{1+\big(\frac{2d}{L}\big)^2}}
\end{equation*} 
the quotient of the two fields is
\begin{equation*}
    \frac{E_{bar}}{E_\infty} = \frac{1}{\sqrt{1+\big(\frac{2d}{L}\big)^2}}.
\end{equation*}
As noted above, the field of the bar approaches that of an infinite
line when $d \ll L$, i.e. for points that are very close to the
bar. Following the aforementioned criterion of admitting a difference
of less than 5\%, for ${E_{bar}}/{E_\infty}$, in this case it must
be verified that $d<0.16L$, i.e., the distance must be less than one
sixth of the length of the bar.

\begin{figure}
    \centerline{\includegraphics[width = 0.7\columnwidth]{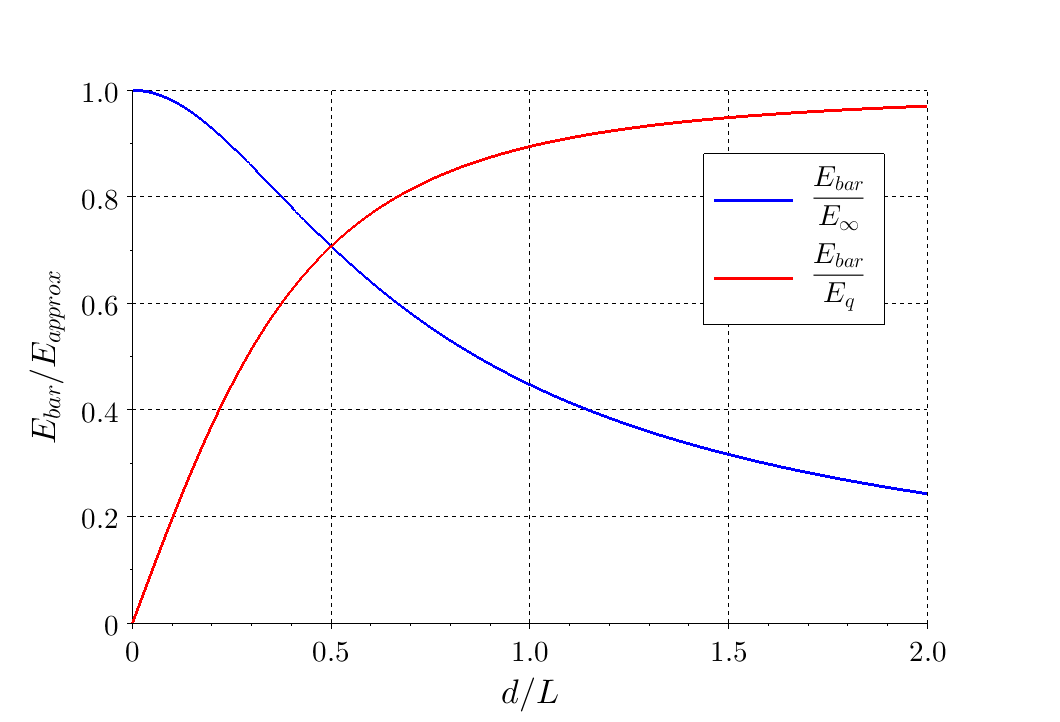}}
    \caption{Comparison of the electric field of a finite charge line
      of length $L$ with the electric fields produced by an infinite
      charge line (blue) and by a point charge (red).}
    \label{fig2}
\end{figure}

A comparison of the electric fields considered above is shown in
Fig.~\ref{fig2}. For small distances, the fields of the finite and
infinite lines  tend to have the same value, whereas for
large distances, the fields of the point charge and the finite line
 have approximately the same value. It is worth noting that
the correction in the electric field in both cases is proportional to
the square of the quotient of the relevant lengths $d$ and $L/2$.

\noindent
\textbf{Using spreadsheets.}
The questions posed about the line of charge, as well as similar
questions about both the electric and magnetic fields, can be answered
numerically using spreadsheets. These spreadsheets are powerful tools
that allow students to easily obtain answers, compare the validity of
different approximations and quickly analyze the dependence with
different parameters \cite{quale2012use,baird2013understanding}.

In order to use feasible values in everyday situations, we assume that
the pen has a length of $L=15cm$ and is charged with $q=10^{-9}C$, a
typical value that can be acquired by rubbing. The different
parameters are entered into the spreadsheet, as well as one step of
distance. Then two columns are drawn for the electric field of the pen
at the observation point: one for the field of a charged particle and
one for the quotient of both quantities.  Based on Fig.~2, we can
see that for this distance it results in $d=24cm$. Therefore, the
electric field of a typical pen at a distance slightly greater than
one and a half times its length behaves approximately like that of a
charged particle.

\noindent
\textbf{Final comments.}
This exercise was proposed to a group of 37 university students in
their first year of General Physics studies, in the form of a
multiple-choice questionnaire with the instruction to answer both
questions without making numerical calculations. The percentage of
correct answers for both questions was about 20\%. In the vast
majority of answers, the maximum distance for the field to resemble
that of a line was overestimated, in many cases by up to two
orders of magnitude. Likewise, the distance required for the field to
resemble that of a point charge was also underestimated by similar
magnitudes. This type of analysis can be performed in the classroom in
just a few minutes, allowing students to analyze different scenarios
and obtain the answer to the validity of different modeling and
approximations numerically. It also allows students to quickly become
familiar with the order of magnitude of electric field values in
different scenarios. The spreadsheet allows them to extend the
analysis to problems without analytical solutions or with more complex
solutions. Moreover, the analysis can be extended to any expression of
electric and magnetic fields, such as a charged disk or a loop through
which current is flowing. We believe that this type of exercise helps
students to develop physical intuition about the idealizations used in
our courses.

\section*{References}

\bibliographystyle{unsrt}

\bibliography{/home/arturo/Dropbox/bibtex/mybib}

\end{document}